\documentclass
[aps,%
12pt,%
final,%
notitlepage,%
oneside,%
onecolumn,%
nobibnotes,%
nofootinbib,%
noshowpacs,%
centertags]%
{revtex4}
\usepackage{amsmath}
\usepackage{epsfig}
\usepackage{endnotes}
\usepackage{graphicx}
\newcommand{\re}{\ref}
\newcommand{\be}{\begin{equation}}
\newcommand{\ee}{\end{equation}}

\newcommand{\la}{\label}
\newcommand{\ber}{\begin{eqnarray}}
\newcommand{\eer}{\end{eqnarray}}

\begin{document}
\title{Some relativistic aspects of nuclear dynamics at the electrodisintegration 
of nuclei}

\author{V.D. Efros\footnote{Electronic address: v.efros@mererand.com}  
    }

\affiliation{National Research Centre 
  "Kurchatov Institute",   Moscow,  Russia}

\begin{abstract}
An approach aimed to extend the applicability range of non-relativistic
microscopic calculations of electronuclear response functions 
is reviewed.
In the quasielastic peak region the calculations agree with experiment at momentum transfers up to about
0.4 GeV/$c$ while at higher momentum transfers, in the region about 0.4 - 1 GeV/$c$, a disagreement is seen. 
In view of this, to calculate the response functions a reference frame was introduced where dynamics relativistic
corrections are small,   and the results pertaining to it
were transformed exactly to the laboratory reference frame.
This proved to remove the major part of the disagreement
with experiment. All leading order relativistic corrections to the transition charge operator and to the one--body part of
the transition current operator were taken into account in the calculations. Furthermore, a particular 
model to determine the kinematics inputs of the non--relativistic
calculations was suggested. This model provides the correct relativistic relationship
between the reaction final--state energy and the momenta of the knocked--out nucleon and the residual system.
The above mentioned choice of  a reference frame in conjunction with this model  has led to an even
better agreement with experiment. 
\end{abstract}

\bigskip


\maketitle

\section{introduction}

At present the nonrelativistic dynamics framework is the only practical one at performing microscopic few-- or many--nucleon
calculations. Last years the predictive power of such calculations increased  due to the progress in the effective field theory approach
to nuclear forces  providing 
a three--nucleon force along with a two--nucleon force. Naturally, it is desirable to test the theory in a wider range of
momentum and energy transfers. However, if one speaks of electrodisintegration reactions in the quasielastic kinematics,
final states of a system may be described nonrelativistically only in a rather limited range of transferred momenta.
Here an approach  making possible the considerable extension of the applicability range of nonrelativistic calculations
is reviewed in short.
The presentation is based on the work done by W. Leidemann, G. Orlandini, E.L. Tomusiak, and the author.

\section{response functions}
In the one photon exchange approximation the inclusive electron scattering
cross section in the laboratory (LAB) frame is given by  
\begin{eqnarray} \label{X}
\frac{d^2\sigma}{d\Omega\,d\omega}\ =\ \sigma_{\rm Mott}\ \bigg[\frac{(q^2-\omega^2)^2}{q^4}\,
R_L(q,\omega)+ \left(\frac{q^2-\omega^2}{2q^2}+\tan^2\frac{\theta}{2}\right)  
\, R_T(q,\omega)\bigg] \,,\la{crs}
\end{eqnarray}
where $R_L$ and $R_T$   
are the longitudinal and transverse response 
functions, respectively.  
The electron variables are denoted by $\omega$ 
(energy transfer), $q$ (momentum transfer), and $\theta$ (scattering angle). All these quantities
pertain to the LAB frame.

Let $\Psi_i$ and $\Psi_f$ be the eigenstates of a nuclear Hamiltonian with energies $E_i$ and $E_f$ and total momenta 
${\bf P}_i$ and ${\bf P}_f$ where $i$ stands for the initial state and $f$ stands for final states of a reaction.
Besides ${\bf P}_f$, the set $f$ includes additional asymptotic quantum numbers that will be denoted by $\tau_f$. 
Let us use the volume element $df=d{\bf P}_fd\tau_f$. The states $\Psi_i$ and $\Psi_f$ are assumed to be normalized as
follows,
\[\langle\Psi_i|\Psi_{i'}\rangle=\delta({\bf P}_i-{\bf P}_{i'}),\qquad
\langle\Psi_f|\Psi_{f'}\rangle=\delta(f-f')\equiv\delta({\bf P}_f-{\bf P}_{f'})\delta(\tau_f-\tau_{f'}).\]

Let $Q({\bf q},\omega)$ and ${\bf J}_t({\bf q},\omega)$ be the nuclear charge operator and the transverse component of the nuclear
current operator ${\bf J}$, ${\bf J}_t={\bf J}-{\hat{\bf q}}({\hat{\bf q}}\cdot{\bf J})$. 
Their matrix elements between the states  $\Psi_i$ and $\Psi_f$ are proportional to $\delta({\bf P}_f-{\bf P}_{i}-{\bf q})$,
\be\langle\Psi_{f}|Q({\bf q},\omega)|\Psi_{i}\rangle=\delta({\bf P}_f-{\bf P}_{i}-{\bf q})Q_{fi},\qquad
\langle\Psi_{f}|{\bf J}_t({\bf q},\omega)|\Psi_{i}\rangle=\delta({\bf P}_f-{\bf P}_{i}-{\bf q})({\bf J}_t)_{fi}.\la{me}\ee 
(${\bf P}_i=0$ in the LAB frame case considered.)
 
The response functions entering Eq. (\re{crs}) are expressed in terms of the on--shell matrix elements from Eqs.~(\re{me}), 
\be R_L(q,\omega)={\overline\sum}_{m_i}\sum\!\!\!\!\!\!\!\!\int d\tau_f\, Q_{if}^\dag Q_{fi}\delta(E_f-E_i-\omega),\la{resp}\ee
\be R_T(q,\omega)={\overline\sum}_{m_i}\sum\!\!\!\!\!\!\!\!\int d\tau_f\, ({\bf J}_t^\dag)_{if}\cdot ({\bf J}_t)_{fi}\delta(E_f-E_i-\omega).
\la{resp1}\ee
Here averagings over the projections of the initial--state
spin are performed and summations plus integrations over $\tau_f$ are present. 

In the nonrelativistic approximation one has $E_{i,f}=P^2_{i,f}/(2M_T)+\epsilon_{i,f}$, where $M_0$ is the mass of a nucleus and 
$\epsilon_{i,f}$ are the energies of internal motion. One also has a representation 
\mbox{$|\Psi_{i,f}\rangle=|{\bf P}_{i,f}\rangle|\psi_{i,f}\rangle$}
where the  states $\psi_{i,f}$ being internal motion states are independent of the quantum numbers ${\bf P}_{i,f}$. 
Let us define the operators $\hat Q$ and ${\hat{\bf J}}_t$ acting in the space
of internal motion states, 
\[{\hat Q}\delta({\bf P}_f-{\bf P}_{i}-{\bf q})=\langle{\bf P}_f|Q|{\bf P}_i\rangle,\qquad
{\hat{\bf J}}_t\delta({\bf P}_f-{\bf P}_{i}-{\bf q})=\langle{\bf P}_f|{\bf J}_t|{\bf P}_i\rangle.\]
Then in accordance with Eqs. (\re{me}) the response functions become
\be R_L(q,\omega)={\overline\sum}_{m_i}\sum\!\!\!\!\!\!\!\!\int d\tau_f\, 
|\langle\psi_f|{\hat Q}(q,\omega)|\psi_i\rangle|^2\delta(\epsilon_f-\epsilon_i+P_f^2/(2M_0)-\omega),\la{nresp}\ee
\be R_T(q,\omega)={\overline\sum}_{m_i}\sum\!\!\!\!\!\!\!\!\int d\tau_f\, 
\langle\psi_i|{\hat{\bf J}}_t^\dag(q,\omega)|\psi_f\rangle\cdot\langle\psi_f|{\hat{\bf J}}_t(q,\omega)|\psi_i\rangle
\delta(\epsilon_f-\epsilon_i+P_f^2/(2M_0)-\omega).\la{nresp1}\ee

\section{nonrelativistic dynamics results}

Response functions of trinucleons were studied.
The nuclear Hamiltonians employed in the calculations included two--body and three--body realistic nuclear interactions
plus the Coulomb force. 
Dependence on  the version of a realistic nuclear force   proved to be weak. 
Most of the results
were obtained with the Argonne V18 two--nucleon interaction \cite{wir} and the Urbana IX three--nucleon interaction \cite{pud}. 
These results
will be presented below. 

In the longitudinal response case the
customary one--body charge transition operator was used in the calculations. It consists of
the nonrelativistic operator and the relativistic corrections of the leading $M_N^{-2}$ order, $M_N$ being the nucleon mass. 
  These are the Darwin--Foldy
and spin--orbit corrections. 
These corrections were accounted for at obtaining the nonrelativistic dynamics results presented below.

The transition operator used in the transverse response case included the nonrelativistic one--body current
consisting of the spin current and convection current plus the usual pion and rho--meson exchange two--body currents. 
The calculations
are described in \cite{04} and in \cite{08}  in the longitudinal case and in the transverse case, respectively, where details 
of the formulation may also be found.

At not low energies it would be very difficult to obtain reaction final states entering Eqs. (\re{nresp}) and (\re{nresp1}) as well as 
to perform the summations and integrations over infinite sets of states there. All this was 
avoided with the help of the method of integral transforms.
The Lorentz transform was employed. The approach is reviewed e.g. in \cite{rl}. A concise review of the formalism can be found,
e.g., in \cite{e12}. The approach is applicable directly when
the operators in Eqs. (\re{nresp}) are $\omega$--independent. Actually they include only numerical factors depending on $\omega$ 
which
may be divided out. A more general case is considered below. 

The numerical results obtained may be considered as being accurate for the present purposes.
This conclusion follows from the studies of convergence trends of the calculations. 
Some of these and similar results were also confirmed with other
methods \cite{gl1,gl2} or other ways to solve arising dynamics equations \cite{bar}.

A review on nonrelativistic studies by other authors of $A$=3 electronuclear reactions is provided in \cite{gl1}.

Let us consider some of the results obtained. 
In Fig.~1 the calculated transverse response functions \cite{08} along with experimental data \cite{mar,dow,car} are shown. 
The agreement with experiment that is observed at $q=250$ MeV/$c$ subsequently deteriorates as $q$ increases. In Fig.~2 the calculated
longitudinal response functions \cite{05} at higher $q$ values are presented along with the data \cite{mar,dow,car}. The nonrelativistic
results we discuss here are shown with the dotted line in the figure. A sharp disagreement is seen also in this case.

\section{privileged reference frame}

In addition to $R_L$ and $R_T$ of Eqs. (\re{resp}) and (\re{resp1}), one may define related responses $R_L^{\rm fr}$ and $R_T^{\rm fr}$.
They are given by the expressions of  Eqs. (\re{resp}) and (\re{resp1}) form with the replacement  
of all the quantities entering there with these quantities but pertaining to another  reference frame. 
Here a class of reference frames is considered which are obtained via boosting 
the LAB frame along {\bf q}. The laboratory responses $R_L$ and $R_T$ can be expressed in terms of 
such $R_L^{\rm fr}$ and $R_T^{\rm fr}$
with the help of
the relationships 
\begin{equation}
R_L(q,\omega) = \frac{q^2}{(q_{\rm fr})^2}{\frac {E_i^{\rm fr}} {M_0}}R_L^{\rm fr}(q_{\rm fr},\omega_{\rm fr}),\qquad 
R_T(q,\omega) = {\frac {E_i^{\rm fr}} {M_0}}R_T^{\rm fr}(q_{\rm fr},\omega_{\rm fr}).\la{relat}
\end{equation}
Here $q_{\rm fr}$, $\omega_{\rm fr}$, and $E_i^{\rm fr}$ are the corresponding quantities pertaining
to a reference frame considered. One has 
\be q_{\rm fr}=\gamma(q-\beta\omega),\quad \omega_{\rm fr}=\gamma(\omega-\beta q),\quad P_i^{\rm fr}=-\beta\gamma M_0,\quad 
E_i^{\rm fr}=\gamma M_0,
\la{kin}\ee
where $\beta c$  is the velocity of a reference frame, $ \gamma=(1-\beta^2)^{-1/2}$,  and $P_i^{\rm fr}$ is
 the initial state momentum in this reference frame.   
The origin of the factor $q^2/(q_{\rm fr})^2$ in (\re{relat}) is shown in \cite{BA1},  
see also, e.g., \cite{DG}. The factor $E_i^{\rm fr}/M_0$ arises since
we adopt the usual normalization of a state of a nucleus to
unity instead of its covariant normalization.

In a genuine relativistic theory any $R_L^{\rm fr}$ and $R_T^{\rm fr}$ from Eqs. (\re{relat}) 
would lead to the same $R_L$ and $R_T$. This, of course, is not the case if $R_L^{\rm fr}$ and $R_T^{\rm fr}$ 
are calculated in the nonrelativistic approximation. 
The responses $R_L$  obtained from $R_L^{\rm fr}$ that are calculated
nonrelativistically are shown in Fig.~2. The anti--lab (AL)
frame, \mbox{${\bf P}_i^{AL}=-{\bf q}_{AL}$}, \mbox{${\bf P}_f^{AL}=0$}, the Breit (B) frame, \mbox{${\bf P}_i^{B}=-{\bf q}_{B}/2$}, 
\mbox{${\bf P}_f^{B}={\bf q}_{B}/2$}, and one more reference frame (ANB) described below were employed. 

The nonrelativistic expressions for the responses $R_L^{\rm fr}$ and $R_T^{\rm fr}$ are of Eqs.~(\re{nresp}) and 
(\re{nresp1}) form with the replacement $P_f^2/(2M_0)$ by $P_f^2/(2M_0)-P_i^2/(2M_0)$ in the $\delta$--function argument. 
The calculation proceeds in the same way as in the LAB frame case. 

In \cite{05} a reference frame aimed to minimize relativistic effects at use of Eqs. (\re{relat}) was introduced.
It was called active--nucleon Breit (ANB) frame. Let us denote $q_{ANB}$ the momentum transfer from the
electron to the nucleus in this frame. The frame is characterized by the fact that in the initial state  
the nucleus has the momentum  $-Aq_{ANB}/2$. At high $q$
values, nucleon momenta in the initial state are thus 
about $-q_{ANB}/2$. In the final state the momentum of
 the active nucleon becomes 
about $q_{ANB}/2$ in quasifree kinematics, while the momentum of each of the other
nucleons remains at about $-q_{ANB}/2$ value. Thus, the typical initial--
and final--state nucleon momenta are restricted to magnitudes
of about $q_{ANB}/2\simeq q/2$ in the ANB reference frame, while,
say, in the laboratory frame nucleon momenta up to $q$ are
present. Correspondingly, the typical relativistic correction to the nucleon
kinetic energy is four times smaller in the ANB frame. 

Furthermore, it also follows from the preceding that
the energy transfer $\omega_{ANB}$ in the ANB reference frame is
zero at the quasielastic peak, and this applies to both the
relativistic and nonrelativistic case. Therefore, even when
one treats the nucleus nonrelativistically the peak remains at
the same position as in the relativistic case. This contrasts
with description of the process in the laboratory reference
frame, where positions of the peak in the relativistic and
nonrelativistic cases would differ considerably. Hence
one expects that performing nonrelativistic calculations in the
quasielastic region in the ANB frame minimizes errors owing
to kinematic relativistic effects. 

Combining Eqs. (\re{kin}), in the ANB frame case one gets 
$\beta=q[2(M_0/A)+\omega]^{-1}\simeq q/(2M_N)$.
(And when substituting 
\mbox{$\omega= [(M_0 /A)^2 +q^2]^{1/2} − M_0 /A$} in the expression (\re{kin}) for $\omega_{ANB}$ one gets $\omega_{ANB} = 0$. 
This is in
agreement with the above statement saying that at the
quasielastic peak $\omega_{ANB} = 0$.)

It is seen from Fig.~2 that, indeed, use of the ANB frame removes the disagreement
with experiment at $q=500$ and 600 MeV/$c$. Comparison with the data at use of the ANB frame is even better if
a contemporary proton form factor in place of the dipole form factor is used \cite{05}. In particular, 
the calculation becomes then closer to the data also at $q=700$ MeV/$c$. In addition, one may note that the 
$q=700$ MeV/$c$ data  are obtained \cite{car} via averaging the data of various groups according to an authors'
judgment.

While differences in Fig.~2 between
the laboratory response $R_L$ calculated directly and $R_L$ calculated from $R_L^{AL}$ and $R_L^{B}$ are considerable 
 such differences
tend to zero at given $q$ and $\omega$ values as the number of nucleons $A$ increases. This is because  ($q_{AL},\omega_{AL}$)
and ($q_B,\omega_B$) tend to ($q,\omega$) at this condition. But ($q_{ANB},\omega_{ANB}$)
are substantially different from ($q,\omega$) at any $A$.\footnote{The ANB frame is thus drastically different from the customary
Breit frame. It is the Breit frame with respect to a constituent of a system while the standard Breit frame is that one with
respect to a whole system consisting of constituents. Correspondingly, the ANB frame makes possible to diminish errors due to relativistic
effects at electrodisintegration processes in quasi elastic kinematics while the Breit frame is useful for this purpose 
in the case of the elastic scattering processes. 
The following comment may also be provided. At the 2013 Jerusalem Conference one of the participants
has suggested that an  approach similar to that reviewed here was developed in the paper by Miller and Thomas, PRC {\bf 56}, 2329
(1997). 
That suggestion was in fact erroneous
and this was admitted subsequently by that person himself.  In the mentioned paper the
standard results obtained with the help of the standard Breit frame for the elastic scattering process
in the cloudy bag model of the nucleon  were merely discussed. The results 
reviewed here have no intersections with those of the mentioned paper.}
 
The ANB frame  was employed also in the calculations of the transverse response \cite{11,10}. 
In addition, relativistic corrections 
to the one--body current operator were obtained and accounted for in these calculations. 
This was done proceeding from the expressions for the matrix elements of 
one--body current of the form $\langle{\bf p}_f |{\bf J}|{\bf p}_i\rangle$ listed 
in \cite{ritz}.                                                     
The operator obtained \cite{10} leads to these expressions and includes all the terms of the orders $M_N^{-1}$ and $M_N^{-3}$.
At calculating $R_T$ this operator was used in the form of a multipole expansion \cite{10}.  In Fig.~3 the effect of relativistic
corrections to the current is shown in the left--hand panels. This effect is considerable. The comparison with experimental data
\cite{mar,dow,car} is shown in the right--hand panels with the dash--dotted line. It is seen that 
use of the ANB frame and accounting for the relativistic
corrections to the current operator removes the most part of the disagreement with experiment seen in Fig.~1. 
(The deviation of the theoretical results from experiment at right wings of the spectra is due to pion production not
included in the theory formulation.) 

Note that use of the special reference frame was of help due to the fact that, in terms of the LAB frame, there exists only
one fast particle in the final state.     

\section{relativistic two--fragment model}

In general, the calculation we consider may be performed nonrelativistically if 
momenta of all the nucleons  are sufficiently small in 
final states of a reaction. When the ANB frame is employed this condition reads as $(q/2M_N)^2\ll1$. 
However, if $q$ would be so large that $q^2/(2M_N)\gg V_N$, $V_N$ being the typical magnitude of the interaction of a knocked--out nucleon 
with a residual subsystem, then this interaction could be neglected. The plane--wave approximation then could be
used to describe the center--of--mass motions of the two fragments. In this case, taking into account relativistic effects
 would merely mean the
use of correct relativistic momenta of the fragments. 

In fact it is known that the mentioned plane--wave approximation is not applicable quantitatively at the $q$ values of interest.
Still, a model in this direction can be constructed which provides a partial accounting for relativistic effects.
This was done \cite{05} in the spirit of the work  on 
electrodisintegration of the deuteron \cite{ar}, where the
relative momentum of outgoing nucleons is determined from a given total energy in the
relativistically correct way, and the relative motion energy that is used as input
to the nonrelativistic  calculation is obtained from that momentum by the
usual  nonrelativistic relation. Usually such a procedure is also applied at
constructing $NN$ potential models.

The model is as follows. Let us denote the momenta of the knocked--out nucleon and that of the
residual nucleus by ${\bf p}_N^{\rm fr}$ and ${\bf p}_X^{\rm fr}$, respectively. Let us express them in terms 
of the total momentum ${\bf P}_f^N$ and the nonrelativistic relative--motion momentum 
\mbox{${\bf p}_{\rm fr}=\mu({\bf p}_N^{\rm fr}/M-{\bf p}_X^{\rm fr}/M_X)$}, where $M_X$ is the mass
of the residual nucleus and $\mu$ is the $N-X$ reduced mass.  (Use of the above 
expression in the present context is not an approximation.) 
 The value of ${\bf p}_{\rm fr}$ can be obtained from the relativistically 
correct kinematical relation
\begin{equation}
E_f^{\rm fr}\ =\ \sqrt{M^2+[{\bf p}_{\rm fr}+(\mu/M_X){\bf P}_f^{\rm fr}]^2}+
\sqrt{M_X^2+[{\bf p}_{\rm fr}-(\mu/M){\bf P}_f^{\rm fr}]^2}.\la{omegafr}
\end{equation}
Then the final--state relative energy 
to use in the nonrelativistic calculations is taken to be
\begin{equation}\label{enr}
\epsilon_f^{\rm fr}=p_{\rm fr}^2/(2\mu)\,.
\end{equation}
It is to be noted that in order to solve Eq. (\ref{omegafr}) for $p_{\rm fr}$ one needs to know its direction.
For the class of reference frames we consider the momentum ${\bf P}_f^{\rm fr}$
is directed along ${\bf q}$. And since we are mainly interested 
in the region of the quasielastic peak we can safely assume that 
${\bf p}_{\rm fr}$ is also directed along ${\bf q}$
(e.g., ${\bf p}_{LAB}\simeq(\mu/M){\bf q}$.) 

For the calculation purpose,
the response functions (\re{resp}) and (\re{resp1}) are to be presented in a form like the nonrelativistic one.
The argument of the energy--conservation $\delta$--function is transformed as follows. One sets $E_f^{\rm fr}=F(\epsilon_f^{\rm fr})$
where $\epsilon_f^{\rm fr}$ is the nonrelativistic energy (\re{enr}). One may write
\begin{equation}
\delta\left(E_f^{\rm fr}-E_i^{\rm fr}-\omega_{\rm fr}\right)\,
=\left(\frac{\partial F}{\partial \epsilon_f^{\rm fr}}\right)^{-1}
\delta\left(\epsilon_f^{\rm fr}-f(q_{\rm fr},\omega_{\rm fr})\right),
\end{equation}
with 
\begin{equation}
\left(\frac{\partial F}{\partial \epsilon_f^{\rm fr}}\right)^{-1}=\frac{p_{\rm fr}}{\mu}
\left(\frac{\partial E_f}{\partial p_{\rm fr}}\right)^{-1}\,.
\end{equation}  
The quantity  $f(q_{\rm fr},\omega_{\rm fr})$ is obtained via the
transformation of the equality $E_f^{\rm fr}=\omega_{\rm fr}+E_i^{\rm fr}$ to the form $\epsilon_f^{\rm fr}=f(q_{\rm fr},\omega_{\rm fr})$. 
Eqs. (\re{omegafr})
and (\re{enr}) are used for this purpose. 

Furthermore, reaction final states are to be taken in the nonrelativistic form that corresponds to Eqs. (\re{nresp}) and (\re{nresp1}). 
The corresponding nonrelativistic energies  of final states are related to their asymptotic momenta as in Eq. (\re{enr}). 
Within the above--mentioned plane--wave approximation such a "nonrelativistic"\,  form would be the exact one.
In general, such a form arises in the  
nonrelativistic approximation. This approximation is the most acceptable when applied in the ANB 
reference frame case. 
As a result, instead of Eqs. (\re{nresp}) and (\re{nresp1}) one gets 
\be R_L(q_{\rm fr},\omega_{\rm fr})=\frac{p_{\rm fr}}{\mu}
\left(\frac{\partial E_f}{\partial p_{\rm fr}}\right)^{-1}{\overline\sum}_{m_i}\sum\!\!\!\!\!\!\!\!\int d\rho_f\, 
|\langle\psi_f|{\hat Q}(q_{\rm fr},\omega_{\rm fr})|\psi_i\rangle|^2\delta\left(\epsilon_f^{\rm fr}-f(q_{\rm fr},\omega_{\rm fr})\right)
\la{nresp2}\ee  
and a similar expression in the transverse case.                                    

To calculate the response (\re{nresp2}) with the help of the above--mentioned method of integral transforms one first considers
$q_{\rm fr}$, $\omega_{\rm fr}$ and $f$ as independent variables and calculates the subsidiary response  
${\bar R}_L(q_{\rm fr},\omega_{\rm fr},f)$
at fixed $q_{\rm fr}$ and $\omega_{\rm fr}$ values. 
Then the cut \mbox{$R_L(q_{\rm fr},\omega_{\rm fr})\equiv{\bar R}_L(q_{\rm fr},\omega_{\rm fr},f(q_{\rm fr},\omega_{\rm fr}))$} is
performed. Such a procedure was suggested in \cite{rei}.

In the right--hand panels of Fig.~3 the results of the application of the above described two--fragment model in conjunction with the
ANB frame ("fr"=$ANB$ in (\re{nresp2})) are shown with the solid line. 
It is seen that the relativization performed according to this model improves the comparison with experiment.

The comparison with experiment became even better when the $\Delta$--isobar configuration degrees of freedom 
in the description of the nuclear system were approximately
taken into account  \cite{yuan}.

Applications of the two--fragment model in case of reference frames different from the ANB frame are considered in \cite{05,11}.
  
The support from RFBR, grant \mbox{13--02--01139}, and from the S.T. Belyaev scientific school grant \mbox{NS--215.2012.2} is acknowledged.

\pagebreak

\centerline{FIGURE CAPTIONS}

\bigskip

{\bf Fig. 1}. Comparison 
of theoretical and experimental $R_T$ at $q =
250$, 400, and 500 MeV/$c$. Theoretical $R_T$ with contributions of
one--body (dotted line) and one--body + two--body transition operators (solid). Experimental data are from \cite{mar}
(triangles), \cite{dow} (circles), and \cite{car} (squares).

\bigskip

{\bf Fig. 2}. Longitudinal response functions calculated in various reference frames.
Experimental data are from \cite{mar} (squares), \cite{dow} (triangles), and \cite{car} (circles).

\bigskip

{\bf Fig. 3}. $R_T(q,\omega)$ at $q=500,$ 600, and 700 MeV/$c$ from the ANB frame calculation. 
Left--hand panels: results without use of two--fragment model with nonrelativistic 
one--body current (dotted line), relativistic one--body current (dashed line), and
relativistic one--body current + MEC (dash--dotted line). Right--hand panels: ANB frame
results with relativistic one--body current + MEC (dash--dotted line). The same with use of two--fragment model (solid line).
Experimental data are from \cite{mar} (squares),  \cite{dow} (diamonds),
\cite{car} (circles).

\begin{figure}[ht]
\centerline{\resizebox*{18.cm}{21.cm}{\includegraphics{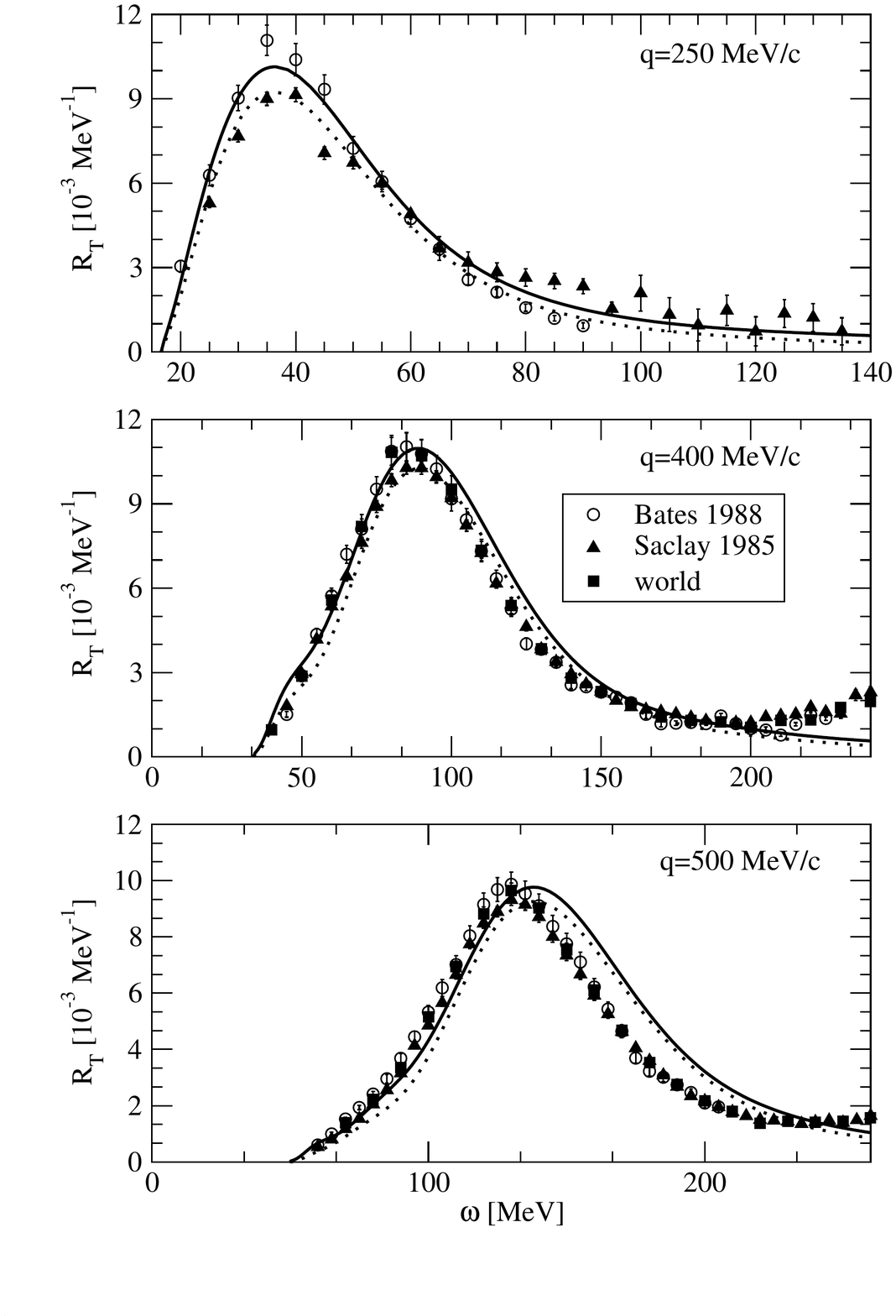}}}
\caption{}
\end{figure}

\begin{figure}[ht]
\centerline{\resizebox*{18.cm}{21.cm}{\includegraphics{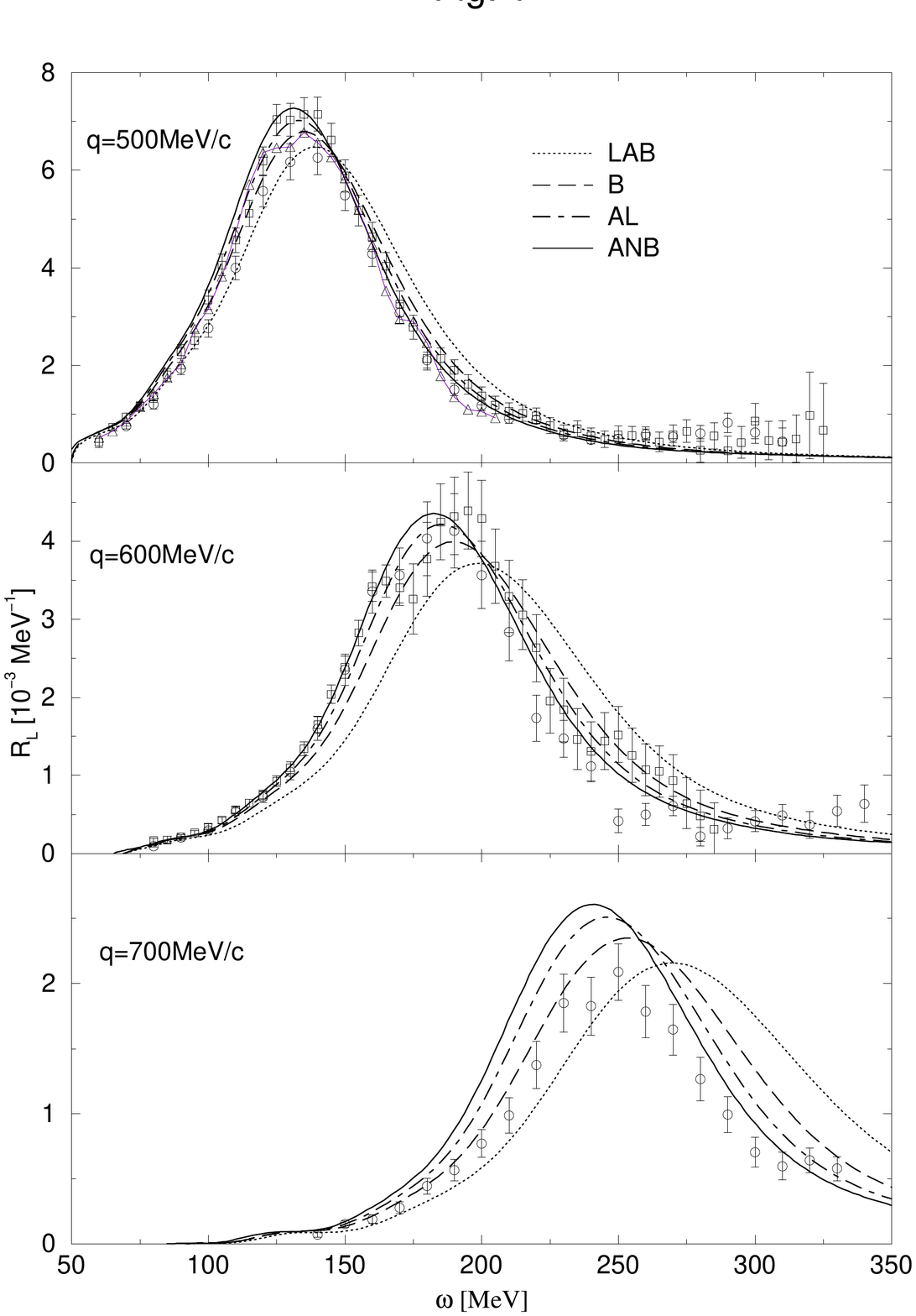}}}
\caption{}
\end{figure}

\begin{figure}[ht]
\centerline{\resizebox*{14.cm}{19.cm}{\includegraphics[angle=0]{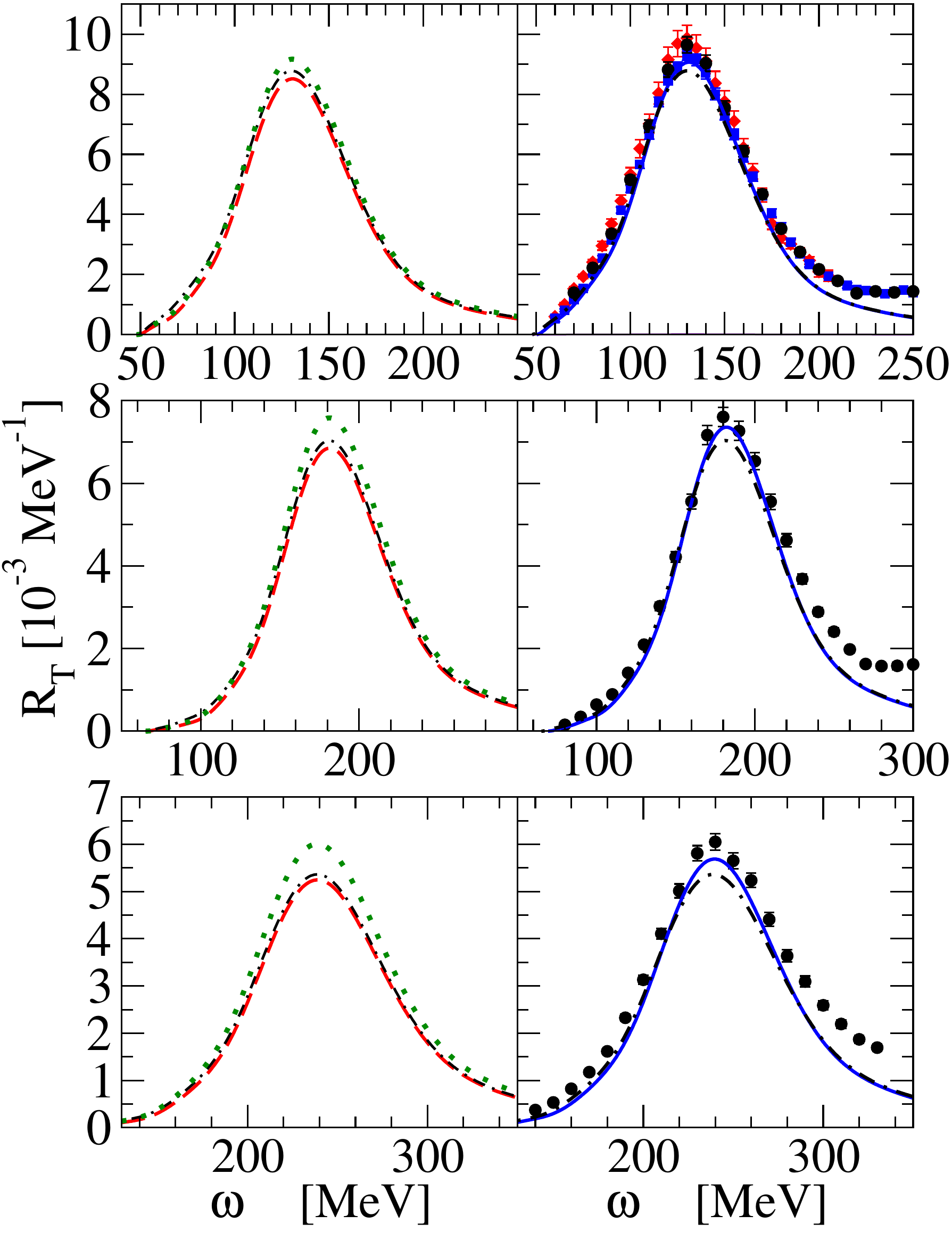}}}
\caption{}
\end{figure}

\end{document}